\documentclass[%
aps,%
12pt,%
final,%
notitlepage,%
oneside,%
onecolumn,%
nobibnotes,%
nofootinbib,%
superscriptaddress,%
noshowpacs,%
showkeys,%
centertags%
]{revtex4}
\usepackage{amsfonts,amssymb,amsbsy}
\usepackage{graphicx,amsfonts}
\usepackage{amsfonts,amssymb,amsbsy}
\usepackage{array}
\usepackage{amsmath}
\usepackage{graphicx,amsfonts}
\usepackage{amsfonts,amssymb,amsbsy}

\begin{document}

\title{Vector meson dominance in $\eta'\rightarrow\pi^0\gamma\gamma$ decay
}

\author{Yaroslav Balytskyi}
\email{BJ@physics.ucla.edu}
\affiliation{Department of Physics and Astronomy, University of California, Los Angeles,  CA 90095-1547, USA
 }

\begin{abstract}

The decay $\eta'\rightarrow\pi^0\gamma\gamma$  is studied  theoretically  in the framework of the Vector Meson Dominance model (VMD). We find theoretically a significant contribution of the interference of $\omega-\rho$ and provide theoretical Dalitz-plots.

Comparison with the experimental results of BES-III \cite{BES-III} is done. We find some tension between our predicted value and the observed result. 

Our calculations can be also checked using the data of GAMS-$4\pi$.

\end{abstract}

\maketitle

\section{Introduction}
\label{intro}

Recently there were obtained a new experimental results \cite{, BES-III, Exp} on the probability of the rare electromagnetic decay 
$
\eta'\rightarrow\pi^0\gamma\gamma
$.

\begin{figure}[h!]
  \includegraphics[width=15cm]{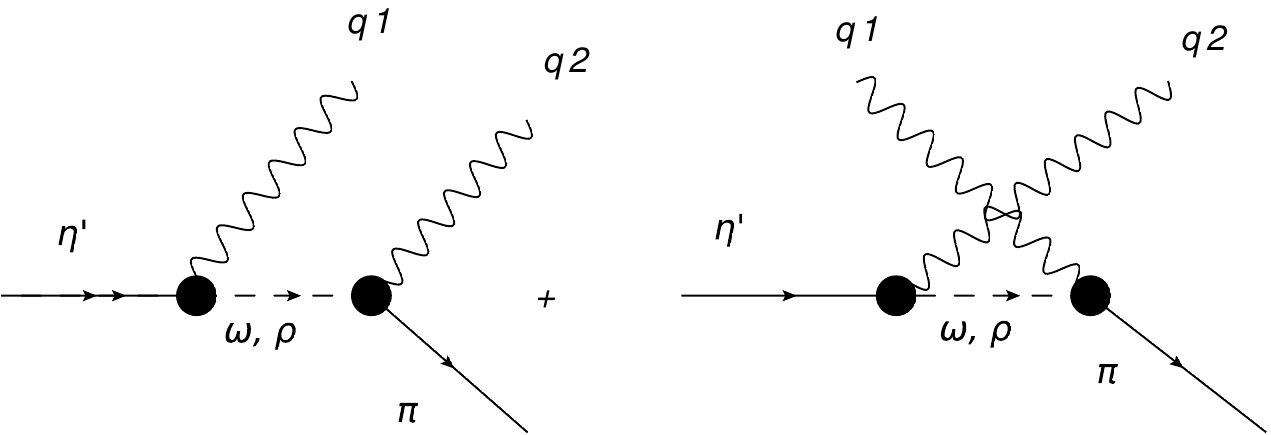}\\
  \caption{Leading order diagrams of $\eta'\rightarrow\pi^0\gamma\gamma$ decay}\label{Diagram}
\end{figure}

From the theory side, the preliminary estimations  in the works \cite{Escribano, Ametller}  on the decay width show that the decay is dominated by the intermediate vector mesons $\omega$ and $\rho$
subsequently decaying into $\pi^0\gamma$,  (Vector Meson Dominance model) \ref{Diagram}.  Contributions both of the chiral loops and linear $\sigma$-terms are suppressed with respect to VMD on the level $\sim 10^{-3}$.

Nevertheless, in these theoretical papers, the role of the interference terms of VMD was not considered. 

Consequently, the decay $\eta'\rightarrow\pi^0\gamma\gamma$ is a unique among similar decays $\eta\rightarrow\pi^0\gamma\gamma$ and $\eta'\rightarrow\eta\gamma\gamma$,  in which the Vector Dominance is manifested in an explicit  and dominant way.

In the general case, the amplitude of the decay \ref{Diagram} is given by
\begin{equation}
A=(\frac{c_{\omega}}{D_{\omega}(t)}+\frac{c_{\rho}}{D_{\rho}(t)})B(q_2)+
(\frac{c_{\omega}}{D_{\omega}(u)}+\frac{c_{\rho}}{D_{\rho}(u)})B(q_1)
\end{equation}
, where $t=(P_{\eta'}-q_1)^2$, $u=(P_{\eta'}-q_2)^2$, $q_1$ and $q_2$ - 4-momenta of outgoing photons, $P_{\eta'}$ - 4-momentum of $\eta'$ - meson, 
$D_{\omega, \rho}(t,u) = (t,u) - im_{\omega, \rho} \Gamma_{\omega, \rho} $ --  
propagator of vector meson (Breit - Wigner function). 
$B(q_{1,2})$ are kinematic coefficients representing the spin structure of the particles.

The constansts of electromagnetic decays are
 $c_{\omega} = g_{\eta'\rightarrow \omega \gamma}\cdot g_{\omega\rightarrow \pi^{0} \gamma},
 c_{\rho}= g_{\eta' \rightarrow \rho \gamma}\cdot g_{\rho \rightarrow \pi^{0} \gamma}$. 

In our work they are determined in 2 ways
   
 \begin{itemize}
 \item From the known decay widths  $\eta' \rightarrow \omega \gamma, \eta' \rightarrow \rho \gamma,
 \omega \rightarrow \pi^{0} \gamma, \rho \rightarrow \pi^{0} \gamma$;
 \item  Using the values of the pseudoscalar mixing angle from our previous work \cite{Mixing} and the work \cite{Phi1}
 \end{itemize}

Estimation made in the work by Escribano \cite{Escribano} predict the decay width to be  
\begin{equation}
\Gamma_{\eta'\rightarrow\pi^0\gamma\gamma}=1.29 keV
\end{equation}
 
The  experimental investigation \cite{PrevExp} of the decay $\eta'\rightarrow\omega\gamma\rightarrow\pi^0\gamma\gamma$ was carried out.

The branching ratio of this decay was found to be

\begin{equation}
BR (\eta' \rightarrow \omega \gamma) = (0.033 \pm 0.004)
\end{equation}

Later, on the experiment GAMS-$4\pi$ the measured branching ratio was reported to be \cite{Exp}
\begin{equation}
BR (\eta' \rightarrow \omega \gamma) = 0.028 \pm 0.003  
\end{equation}

In our work we neglect the contribution of the chiral loops and $\sigma$-terms taking into account only VMD. We consider different values of the coupling constants, first, extracted from the decays which are known. Second, assuming OZI rule is strictly fulfilled and taking the coupling constant from a known mixing angles.

Our theoretical predictions  \ref{Prediction} are the  following

 \begin{itemize}
\item Although the coupling constants of $\omega$ and $\rho$ mesons are approximately the same, since $\omega$ has a much smaller width, it dominates and gives $\approx 75\%$ of the contribution to the integral over Dalitz-plot.
\item The corresponding contribution of $\rho$ - meson is $\approx 5\%$.

\item Nevertheless, the contribution of $\omega-\rho$ interference is big  $\approx 20 \%$. 
\end{itemize}

\section{Theoretical predictions}\label{Prediction}
\subsection{Matrix element}
After the sum over polarization the amplitude squared of matrix element takes the form

\begin{equation}
 M_{fin}  = B(q_{2})B(q_{2})BW(t) + 2B(q_{2})B(q_{1})BW_{int} + B(q_{1})B(q_{1})BW(u)
 \label{matr}
\end{equation}

, where $BW(t,u)$  - amplitude squared  of the Breit-Wigner function, $BW_{int}$ -- interference term constructed from Breit-Wigner functions, and
$B(q_{2})B(q_{2})$, $B(q_{2})B(q_{1})$  and $B(q_{1})B(q_{1})$ - kinematic coefficients obtained after taking a sum over polarization of $\gamma$.

The decay width of vector mesons are taken from \cite{pdg}. Moreover, in this work we consider the decay width of the $\rho$ - meson which depends on the energy.

In terms of 4-momenta the kinematic coefficients are given by

\begin{math}
B(q_2)B(q_2)=
(q_1\cdot q_2)^2[(P_{\eta'}\cdot q_2)^2+((P_{\eta'}\cdot q_2)-m^2_{\eta'})^2] +
2(P_{\eta'}\cdot q_1)(P_{\eta'}\cdot q_2)[(P_{\eta'}\cdot q_1)(P_{\eta'}\cdot q_2)-m^2_{\eta'}(q_1\cdot q_2)]
\end{math}

\begin{math}
B(q_2)B(q_1)=
(q_1\cdot q_2)^2[3(P_{\eta'}\cdot q_1)(P_{\eta'}\cdot q_2)-2m^2_{\eta'}((P_{\eta'}\cdot q_1)+(P_{\eta'}\cdot q_2)-m^2_{\eta'})]+
2(P_{\eta'}\cdot q_1)(P_{\eta'}\cdot q_2)[(P_{\eta'}\cdot q_1)(P\cdot q_2)+(q_1\cdot q_2)((P_{\eta'}\cdot q_1)+(P_{\eta'}\cdot q_2)-2 m^2_{\eta'})]
\end{math}

\begin{math}
B(q_1)B(q_1)=
(q_1,q_2)^2[(P_{\eta'}\cdot q_1)^2+((P_{\eta'}\cdot q_1)-m^2_{\eta'})^2] +
2(P_{\eta'}\cdot q_1)(P_{\eta'}\cdot q_2)[(P_{\eta'}\cdot q_1)(P_{\eta'}\cdot q_2)-m^2_{\eta'}(q_1\cdot q_2)]
\end{math}

, where $P_{\eta'}$ and $p_{\pi}$ - momenta of $\eta'$ and $\pi^0$ respectively, and $q_{1, 2}$ - momenta of outcoming photons.

Now we switch to Dalitz variables which are defined by: $m_{13}=(q_1+p_{\pi^0})^2$ and $m_{23}=(q_2+p_{\pi^0})^2$. 

Taking into account that  $P_{\eta'}=p_{\pi}+q_1+q_2$, we find the products of 4-vectors in terms of Dalitz variables:

\begin{math}
(p_{\pi}\cdot q_1)=\frac{m^2_{13}-m^2_{\pi}}{2}, (p_{\pi}\cdot q_2)=\frac{m^2_{23}-m^2_{\pi}}{2}, (q_1\cdot q_2)=\frac{m^2_{\eta'}+m^2_{\pi}-m^2_{13}-m^2_{23}}{2},
(P_{\eta'}\cdot q_1)=\frac{m^2_{\eta'}-m^2_{23}}{2}, 
(P_{\eta'}\cdot q_2)=\frac{m^2_{\eta'}-m^2_{13}}{2}
\end{math}

Now the kinematic coefficients take the form:

\begin{math}
B(q_2)B(q_2)=\frac{1}{8}\Bigl[(m^2_{\eta'}+m^2_{\pi}-m^2_{13}-m^2_{23})^2(m^4_{\eta'}+m^4_{13})+(m^2_{\eta'}-m^2_{13})(m^2_{\eta'}-m^2_{23})
\bigl(m^2_{13}m^2_{23}+m^2_{\eta'}(m^2_{13}+m^2_{23}-2m^2_{\pi}-m^2_{\eta'})\bigr)\Bigr]
\end{math}

\begin{math}
B(q_2)B(q_1)=\frac{1}{16}(m^2_{\eta'}+m^2_{\pi}-m^2_{13}-m^2_{23})^2 (m^2_{\eta'}(3m^2_{\eta'}+m^2_{13}+m^2_{23})+3m^2_{13}m^2_{23})+\frac{1}{8}(m^2_{\eta'}-m^2_{13})(m^2_{\eta'}-m^2_{23})[(m^2_{13}+m^2_{23})^2+m^2_{13}m^2_{23}-m^4_{\eta'}-m^2_{\pi}(m^2_{13}+m^2_{23}+2m^2_{\eta'})]
\end{math}

\begin{math}
B(q_1)B(q_1)=\frac{1}{8}\Bigl[(m^2_{\eta'}+m^2_{\pi}-m^2_{13}-m^2_{23})^2(m^4_{\eta'}+m^4_{23})+(m^2_{\eta'}-m^2_{13})(m^2_{\eta'}-m^2_{23})
(m^2_{13}m^2_{23}+m^2_{\eta'}(m^2_{13}+m^2_{23}-2m^2_{\pi}-m^2_{\eta'}))\Bigr]
\end{math}

Breit-Wigner functions are given by:

\begin{math}
BW(t, u)=\frac{c^2_{\omega}}{(m^2_{\omega}-m^2_{13,23})^2+m^2_{\omega}\Gamma^2_{\omega}}+
\frac{c^2_{\rho}}{(m^2_{\rho}-m^2_{13,23})^2+m^2_{\rho}\Gamma^2_{\rho}}+
2c_{\omega}c_{\rho}\frac{(m^2_{\omega}-m^2_{13,23})(m^2_{\rho}-m^2_{13,23})+m_{\omega}\Gamma_{\omega}m_{\rho}\Gamma_{\rho}}{((m^2_{\omega}-m^2_{13,23})^2+m^2_{\omega}\Gamma^2_{\omega})((m^2_{\rho}-m^2_{13,23})^2+m^2_{\rho}\Gamma^2_{\rho})}
\end{math}

\begin{math}
BW_{int}=c^2_{\omega}\cdot\frac{(m^2_{\omega}-m^2_{23})(m^2_{\omega}-m^2_{13})+m^2_{\omega}\Gamma_{\omega}^2}{((m^2_{\omega}-m^2_{23})^2+m^2_{\omega}\Gamma^2_{\omega})((m^2_{\omega}-m^2_{13})^2+m^2_{\omega}\Gamma^2_{\omega})}+
c_{\omega}c_{\rho}[\frac{(m^2_{\omega}-m^2_{23})(m^2_{\rho}-m^2_{13})+m_{\omega}\Gamma_{\omega}m_{\rho}\Gamma_{\rho}}{((m^2_{\omega}-m^2_{23})^2+m^2_{\omega}\Gamma^2_{\omega})((m^2_{\rho}-m^2_{13})^2+m^2_{\rho}\Gamma^2_{\rho})}+
\frac{(m^2_{\rho}-m^2_{23})(m^2_{\omega}-m^2_{13})+m_{\rho}\Gamma_{\rho}m_{\omega}\Gamma_{\omega}}{((m^2_{\rho}-m^2_{23})^2+m^2_{\rho}\Gamma^2_{\rho})((m^2_{\omega}-m^2_{13})^2+m^2_{\omega}\Gamma^2_{\omega})}]+
c^2_{\rho}\cdot\frac{(m^2_{\rho}-m^2_{23})(m^2_{\rho}-m^2_{13})+m^2_{\rho}\Gamma_{\rho}^2}{((m^2_{\rho}-m^2_{23})^2+m^2_{\rho}\Gamma^2_{\rho})((m^2_{\rho}-m^2_{13})^2+m^2_{\rho}\Gamma^2_{\rho})}
\end{math}

, where $t$ corresponds to $m_{13}$ and $u$ to $m_{23}$ respectively.

\subsubsection{Interference between $\rho-\omega$}
In order to understand our expression better, we split the matrix element in the form of  
 $\rho-\omega$ contributions
 
 $$
 M_{total}=M_{\omega}+2\cdot M_{\omega-\rho}+M_{\rho}
 $$

\begin{math}
M_{\omega, \rho}= c^2_{\omega, \rho}\cdot [ \frac{B(q_{2})B(q_{2})}{(m^2_{\omega, \rho}-m^2_{13})^2+m^2_{\omega, \rho}\Gamma^2_{\omega, \rho}}+\frac{B(q_{1})B(q_{1})}{(m^2_{\omega, \rho}-m^2_{23})^2+m^2_{\omega, \rho}\Gamma^2_{\omega, \rho}}+2 B(q_{2})B(q_{1})
\cdot\frac{(m^2_{\omega, \rho}-m^2_{23})(m^2_{\omega, \rho}-m^2_{13})+m^2_{\omega, \rho}\Gamma_{\omega, \rho}^2}{((m^2_{\omega, \rho}-m^2_{23})^2+m^2_{\omega, \rho}\Gamma^2_{\omega, \rho})((m^2_{\omega, \rho}-m^2_{13})^2+m^2_{\omega, \rho}\Gamma^2_{\omega, \rho})}]
\end{math}

\begin{math}
M_{\omega-\rho}= c_{\omega}\cdot c_{\rho}\cdot [B(q_{2})B(q_{2})\cdot
(\frac{(m^2_{\omega}-m^2_{13})(m^2_{\rho}-m^2_{13})+m_{\omega}\Gamma_{\omega}m_{\rho}\Gamma_{\rho}}{((m^2_{\omega}-m^2_{13})^2+m^2_{\omega}\Gamma^2_{\omega})((m^2_{\rho}-m^2_{13})^2+m^2_{\rho}\Gamma^2_{\rho})})+
B(q_{1})B(q_{1})\cdot(\frac{(m^2_{\omega}-m^2_{23})(m^2_{\rho}-m^2_{23})+m_{\omega}\Gamma_{\omega}m_{\rho}\Gamma_{\rho}}{((m^2_{\omega}-m^2_{23})^2+m^2_{\omega}\Gamma^2_{\omega})((m^2_{\rho}-m^2_{23})^2+m^2_{\rho}\Gamma^2_{\rho})})+
B(q_2)B(q_1)\cdot(\frac{(m^2_{\omega}-m^2_{23})(m^2_{\rho}-m^2_{13})+m_{\omega}\Gamma_{\omega}m_{\rho}\Gamma_{\rho}}{((m^2_{\omega}-m^2_{23})^2+m^2_{\omega}\Gamma^2_{\omega})((m^2_{\rho}-m^2_{13})^2+m^2_{\rho}\Gamma^2_{\rho})}+
\frac{(m^2_{\rho}-m^2_{23})(m^2_{\omega}-m^2_{13})+m_{\rho}\Gamma_{\rho}m_{\omega}\Gamma_{\omega}}{((m^2_{\rho}-m^2_{23})^2+m^2_{\rho}\Gamma^2_{\rho})((m^2_{\omega}-m^2_{13})^2+m^2_{\omega}\Gamma^2_{\omega})})]
\end{math}

In section \ref{Numerical} we calculate the relative contribution of these terms.

\subsection{Coupling constants}
The coupling constants are not fixed uniquely and are the main source of theoretical uncertainty. 

\subsubsection{Extraction from the known decays}
First, we calculate them from the known decays,  $\omega\rightarrow\pi^0\gamma, \rho\rightarrow\pi^0\gamma, \eta'\rightarrow\omega\gamma, \eta'\rightarrow\rho\gamma$, the details of calculation are provided in \ref{Appendix}.

The decay width $\omega\rightarrow\pi^0\gamma$ is given by

\begin{equation}
\Gamma(\omega\rightarrow\pi^0\gamma)=\frac{1}{3}\cdot g_{\omega\rightarrow\pi^0\gamma}^2 \cdot \frac{(m^2_{\omega}-m^2_{\pi})^3}{32\pi\cdot m^3_{\omega}}
\end{equation}

And the decay constant is $ g_{\omega\rightarrow\pi^0\gamma}^2=0.484168 GeV^{-2}$.

Analogously, for  a decay $\rho\rightarrow\pi^0\gamma$ 

\begin{equation}
\Gamma(\rho\rightarrow\pi^0\gamma)=\frac{1}{3}\cdot g_{\rho\rightarrow\pi^0\gamma}^2 \cdot \frac{(m^2_{\rho}-m^2_{\pi})^3}{32\pi\cdot m^3_{\rho}}
\end{equation}

The corresponding decay constant is $g_{\rho\rightarrow\pi^0\gamma}^2=0.0635057 GeV^{-2}$.

For the $\eta'\rightarrow\omega\gamma$ and $\eta'\rightarrow\rho\gamma$ decays there is no $\frac{1}{3}$ factor since $\eta'$ is spin zero.

\begin{equation}
\Gamma(\eta'\rightarrow\omega\gamma)= g_{\eta'\rightarrow\omega\gamma}^2 \cdot \frac{(m^2_{\eta'}-m^2_{\omega})^3}{32\pi\cdot m^3_{\eta'}}
\end{equation}

\begin{equation}
\Gamma(\eta'\rightarrow\rho\gamma)= g_{\eta'\rightarrow\rho\gamma}^2 \cdot \frac{(m^2_{\eta'}-m^2_{\rho})^3}{32\pi\cdot m^3_{\eta'}}
\end{equation}

The corresponding decay constants are $g_{\eta'\rightarrow\omega\gamma}^2=0.0169841 GeV^{-2}$ and
$g_{\eta'\rightarrow\rho\gamma}^2=0.160804 GeV^{-2}$

Finally, the coupling constants for $\omega$ and $\rho$ mesons are the following:
\begin{equation}
c_{\omega}=g_{\eta'\rightarrow \omega \gamma}\cdot g_{\omega\rightarrow \pi^{0} \gamma}=0.0906816 GeV^{-2}
\end{equation}

\begin{equation}
c_{\rho}= g_{\eta' \rightarrow \rho \gamma}\cdot g_{\rho \rightarrow \pi^{0} \gamma}=0.101054 GeV^{-2}
\end{equation}

As we see, the coupling constants in this approach are only approximately the same, and the discrepancy between them is of order $\sim 10 \%$

\subsubsection{Determination of the coupling constants assuming exact OZI.}

In the limit of an exact OZI rule the coupling constants are connected by isospin factors and the pseudoscalar mixing angle $\phi_P$.

The current is $j^{\mu}=\frac{2}{3}\bar{u}\gamma^{\mu}u-\frac{1}{3}\bar{d}\gamma^{\mu}d=\underbrace{\frac{1}{2}(\bar{u}\gamma^{\mu}u-\bar{d}\gamma^{\mu}d)}_{I=1}+\underbrace{\frac{1}{6}(\bar{u}\gamma^{\mu}u+\bar{d}\gamma^{\mu}d)}_{I=0}.$

Taking into account $I(\eta')=0$, $I(\omega)=0$, $I(\rho)=1, I(\pi^0)=1$. The isospin multiplyers are:

$$I(\eta'\rightarrow\omega\gamma\rightarrow\pi^0\gamma\gamma)=\frac{1}{6}\cdot\frac{1}{2}=\frac{1}{12}$$

$$I(\eta'\rightarrow\rho\gamma\rightarrow\pi^0\gamma\gamma)=\frac{1}{2}\cdot\frac{1}{6}=\frac{1}{12}$$

Consequantly, in this approach the coupling constants are exactly the same:

\begin{math}
c^{OZI}_{\omega}=c^{OZI}_{\rho}=g_{\rho\eta'\gamma}\cdot g_{\rho\pi^0\gamma}=g_{\omega\eta'\gamma}\cdot g_{\omega\pi^0\gamma}=
(\frac{Ge}{\sqrt{2}g})^2\cdot \frac{1}{3}\cdot  Sin[{\phi_P}], 
\end{math}

, where $G=\frac{3 g^2}{4\pi^2 f_{\pi}}, g\approx 4.2, f_{\pi}\approx 93 MeV = 0.093GeV$

$$
(\frac{Ge}{\sqrt{2}g})^2=(\frac{e}{\sqrt{2}g}\frac{3g^2}{4\pi^2 f_{\pi}})^2=\frac{9g^2}{8\pi^3\cdot f_{\pi}^2}\cdot \alpha = 0.540012     GeV^{-2}
$$

Nevertheless, the mixing angle is not uniquely defined.

We derived the mixing angle in our previous work \cite{Mixing} to be $\phi_P=37.4^\circ \pm0.4 ^\circ$.

In \cite{Phi1} the previous results on determination of the mixing angle in different ways are summarized. They provide several values extracted in different ways:

\begin{math}
\phi_P=44.2^\circ\pm 1.4^\circ;
43.2^\circ\pm2.8 ^\circ; 40.7^\circ\pm3.7 ^\circ; 42.7^\circ\pm5.4^\circ; 41.0^\circ\pm3.5^\circ; 41.2^\circ\pm3.7^\circ; 50^\circ\pm26^\circ; 36.5^\circ\pm1.4^\circ; 
42.4^\circ\pm2.0^\circ; 40.2^\circ\pm2.8^\circ
\end{math}

In \ref{Zvedennya} the values of the coupling constants extracted in different ways are summarized.

\begin{table}[h!]\label{Zvedennya}
\caption{Coupling constants}
\begin{center}
\begin{tabular}{|l|c|r|}
\hline
Coupling constant & $c_{\omega,\eta'\rightarrow\pi^0\gamma\gamma}$, $GeV^{-2}$ & $c_{\rho,\eta'\rightarrow\pi^0\gamma\gamma}$, $GeV^{-2}$ \\ \hline
Extracted from the known decays. & 0.090682 & 0.101054 \\  \hline
$\phi_P=37.4^\circ$(from our previous work \cite{Mixing}) & 0.109331 & 0.109331 \\ \hline
$\phi_P=44.2^\circ$ & 0.125493 & 0.125493 \\ \hline
$\phi_P=43.2^\circ$ & 0.123221 & 0.123221 \\ \hline
$\phi_P=40.7^\circ$ & 0.117380 & 0.117380 \\ \hline
$\phi_P=42.7^\circ$ & 0.122072 & 0.122072 \\ \hline
$\phi_P=41.0^\circ$ & 0.118093 & 0.118093 \\ \hline
$\phi_P=41.2^\circ$ & 0.118567 & 0.118567 \\ \hline
$\phi_P=50^\circ$ & 0.137891 & 0.137891 \\ \hline
$\phi_P=36.5^\circ$ & 0.107071 & 0.107071 \\ \hline
$\phi_P=42.4^\circ$ & 0.121377 & 0.121377 \\ \hline
$\phi_P=40.2^\circ$ & 0.116185 & 0.116185 \\ \hline
\end{tabular}
\end{center}
\end{table} 

We see, the coupling constants can vary up to $\sim 30\%$ which can lead to a variation in the predicted decay width up to $\sim 50 \%$.

\subsection{Dynamical width of $\rho$ - meson}

The $\omega$ - meson is quite narrow and its peaks are clearly seen on a Dalitz plot, but $\rho$ - meson is wide, so we have to estimate the corrections due to the dependence of the width of $\rho$ - meson on energy.

The dependence of the width of $\rho$ meson is dictated by the unitarity conditions  \cite{PionLoop}.

Breit-Wigner function in the general form is:
\begin{equation}
f(s)=\frac{1}{s_r-s-i\rho(s)\cdot g}
\end{equation}

For a $p$ -wave $\rho\sim p^3$, for the case of a scalar particle it would be $\rho\sim p^1$.

For the width of  $\rho$ -meson  the following parametrization was used:
\begin{equation}
\lambda[x, y, z] = x^2 + y^2 + z^2 - 2 x y - 2 y z - 2 z x;
\end{equation}
\begin{equation}
p =\frac{\sqrt{\lambda[s, m_{\pi}^2, m_{\pi}^2]}}{2\sqrt{s}}
\end{equation}
\begin{equation}
p_0 =\frac{\sqrt{\lambda[m^2_{\rho}, m_{\pi}^2, m_{\pi}^2]}}{2 m_{\rho}}
\end{equation}
\begin{equation}
R=5 GeV^{-1} --->interaction  length
\end{equation}
\begin{equation}\label{Mikhasenko}
\Gamma^{eff}_{\rho}=\Gamma_{\rho}(\frac{p}{p_0})^3 \frac{m_{\rho}}{\sqrt{s}}\frac{1+p^2_0\cdot R^2}{1+p^2\cdot R^2}
\end{equation}
, where $m_{\pi}$ - the mass of $\pi^0$.

\begin{figure}
  \includegraphics[width=10cm]{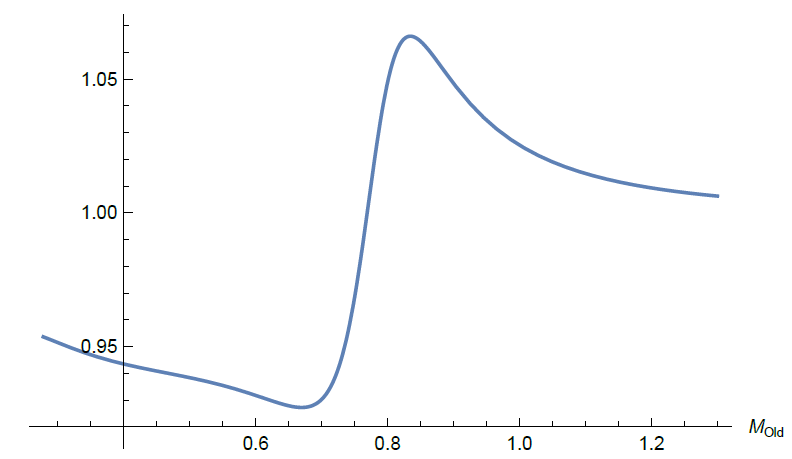}\\
  \caption{The ratio (Breit-Wigner function with a dynamical width of $\rho$-meson)/(Breit-Wigner function with a constant width of $\rho$-meson)}\label{Dynamics}
\end{figure}

We see, the dynamical width of  $\rho$ meson can give only up to $\sim 5\%$ contribution to the Breit-Wigner function \ref{Dynamics}. So, it is irrelevant in our further investigation since the uncertainty of the coupling constant is much larger. 

\subsection{Dalitz plots and decay widths predictions}\label{Numerical}

Now we make theoretical Dalitz - plots of the decay and calculate the decay width. 

The constraints on the boundaries of the Dalitz - plot are the following:
\begin{equation}
\begin{cases}
(m^2_{23})_{max}=(\frac{M^2_{\eta'}+m^2_{\pi}}{2 m_{13}})^2
-(\sqrt{(\frac{m^2_{13}+m^2_{\pi}}{2 m_{13}})^2-m^2_{\pi}}-\frac{M^2_{\eta'}-m^2_{13}}{2m_{13}})^2 \\
(m^2_{23})_{min}=(\frac{M^2_{\eta'}+m^2_{\pi}}{2 m_{13}})^2
-(\sqrt{(\frac{m^2_{13}+m^2_{\pi}}{2 m_{13}})^2-m^2_{\pi}}+\frac{M^2_{\eta'}-m^2_{13}}{2m_{13}})^2
\end{cases}
\end{equation}

The decay width and the expression for 3-particle phase space are the following \cite{SlSp}:

$$d\Gamma=\frac{(2\pi)^4}{2M_{\eta'}}|A|^2 dR_3$$, 

$$dR_3=\frac{1}{2}\cdot\frac{1}{(2\pi)^9}\frac{\pi^2}{4s}dm_{13}^2dm_{23}^2$$.

, where the factor $\frac{1}{2}$ comes from the fact that we have 2 identical $\gamma$-quants in the final state. 

Assuming the initial $\eta'$ to be at rest, we obtain:

$$d\Gamma=\frac{1}{2}\cdot\frac{1}{256\pi^3M_{\eta'}^2}|A|^2dm^2_{13}dm^2_{23}$$

After that, the integral is taken numerically. The points are randomly dropped on the \textbf{squares},  $[m^2_{13}, m^2_{23}]$, since the integral is taken over the squares of Dalitz variables. The points which don't hit the Dalitz - plot are truncated. 

After that the integral is replaced by a finite sum.

\begin{table}[h!]\label{WidthsZvedennya}
\caption{Coupling constants}
\begin{center}
\begin{tabular}{|l|c|r|}
\hline
Coupling constant & The decay width $\Gamma(\eta'\rightarrow\pi^0\gamma\gamma)$, keV\\ \hline
Extracted from the known decays. & 1.7  \\  \hline
$\phi_P=37.4^\circ$ & 2.4  \\ \hline
$\phi_P=44.2^\circ$ & 3.2  \\ \hline
$\phi_P=43.2^\circ$ & 3.0  \\ \hline
$\phi_P=40.7^\circ$ & 2.8  \\ \hline
$\phi_P=42.7^\circ$ & 3.0  \\ \hline
$\phi_P=41.0^\circ$ & 2.8  \\ \hline
$\phi_P=41.2^\circ$ & 2.8  \\ \hline
$\phi_P=50^\circ$ & 3.8  \\ \hline
$\phi_P=36.5^\circ$ & 2.3  \\ \hline
$\phi_P=42.4^\circ$ & 3.0 \\ \hline
$\phi_P=40.2^\circ$ & 2.7 \\ \hline
\end{tabular}
\end{center}
\end{table} 

Consequently, we see, the prediction can significantly vary depending on the coupling constant used:

\begin{equation}
\Gamma(\eta'\rightarrow\pi^0\gamma\gamma) = 1.7 - 3.8 keV
\end{equation}

\begin{figure}
\includegraphics[width=10cm]{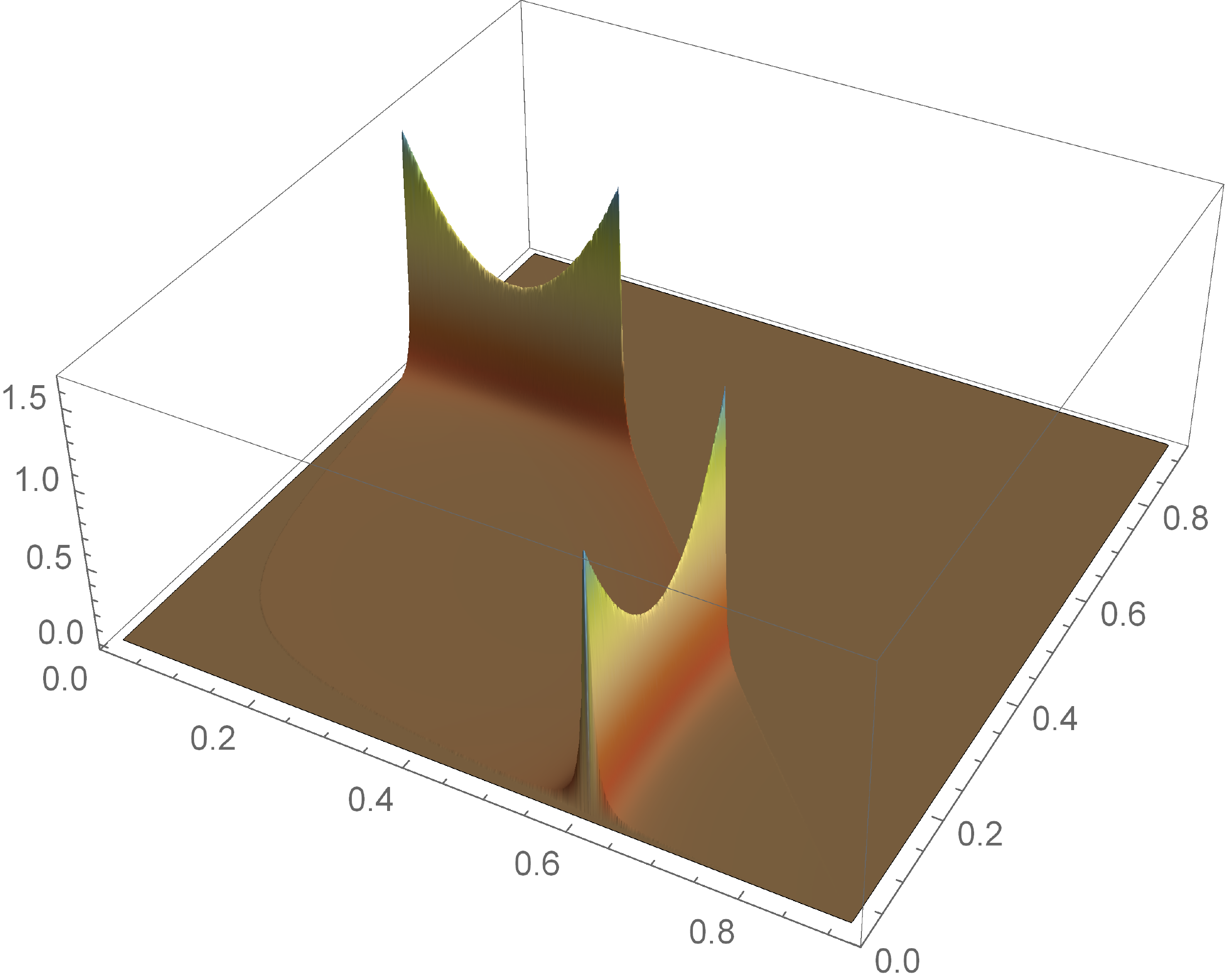}
\caption{Theoretical Dalitz plot with $\phi_P=44.2^\circ$}
\label{DalitzTheoretical}
\end{figure}

\begin{figure}
\includegraphics[width=10cm]{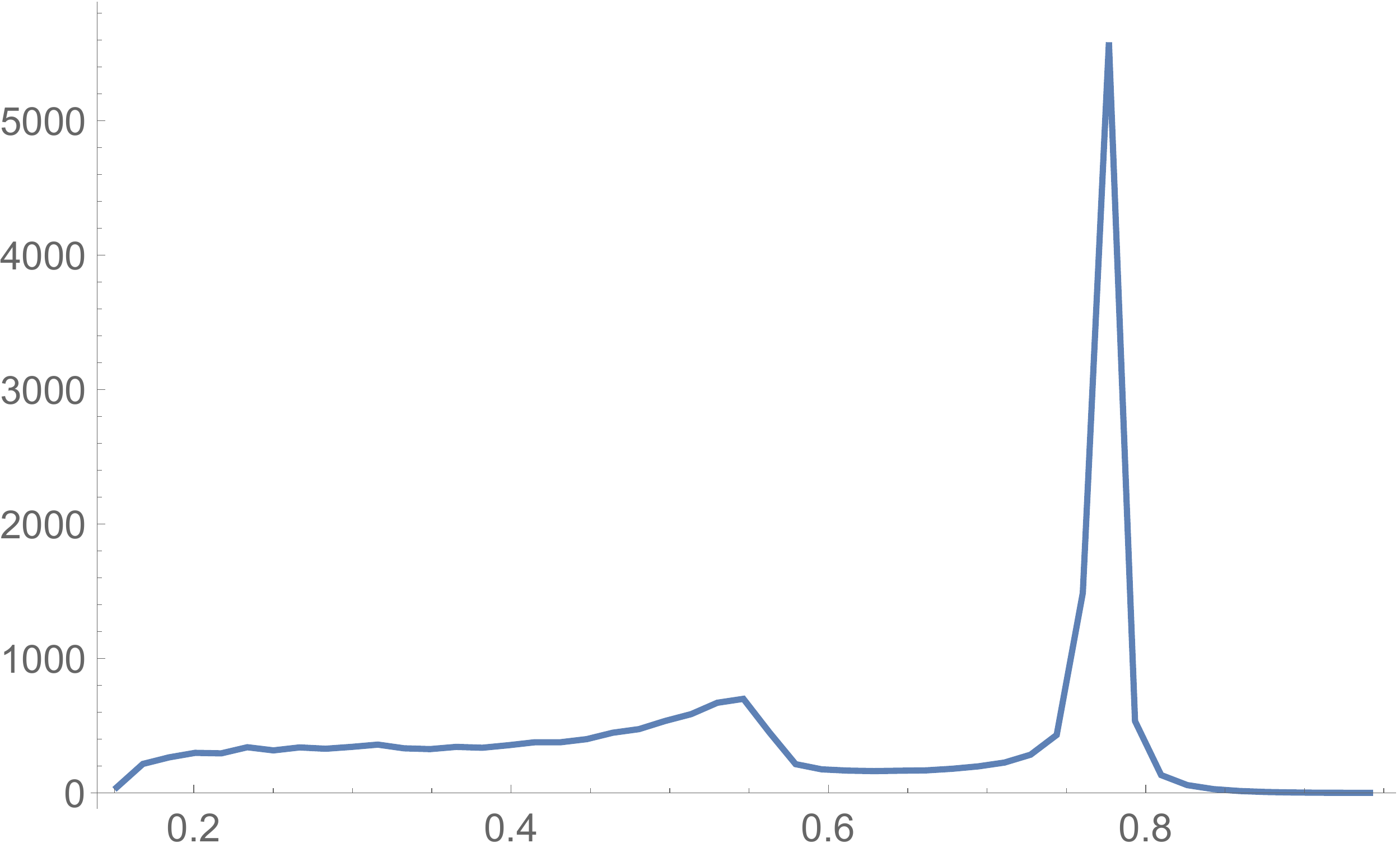}
\caption{Theoretical mass spectrum of $\pi^0\gamma$ with $\phi_P=44.2^\circ$}
\label{MassSpectrumTheoretical}
\end{figure}

The theoretical Dalitz plot and mass spectrum of $\pi^0\gamma$ are shown on the pictures \ref{DalitzTheoretical}, \ref{MassSpectrumTheoretical}.

\begin{figure}
\includegraphics[width=10cm]{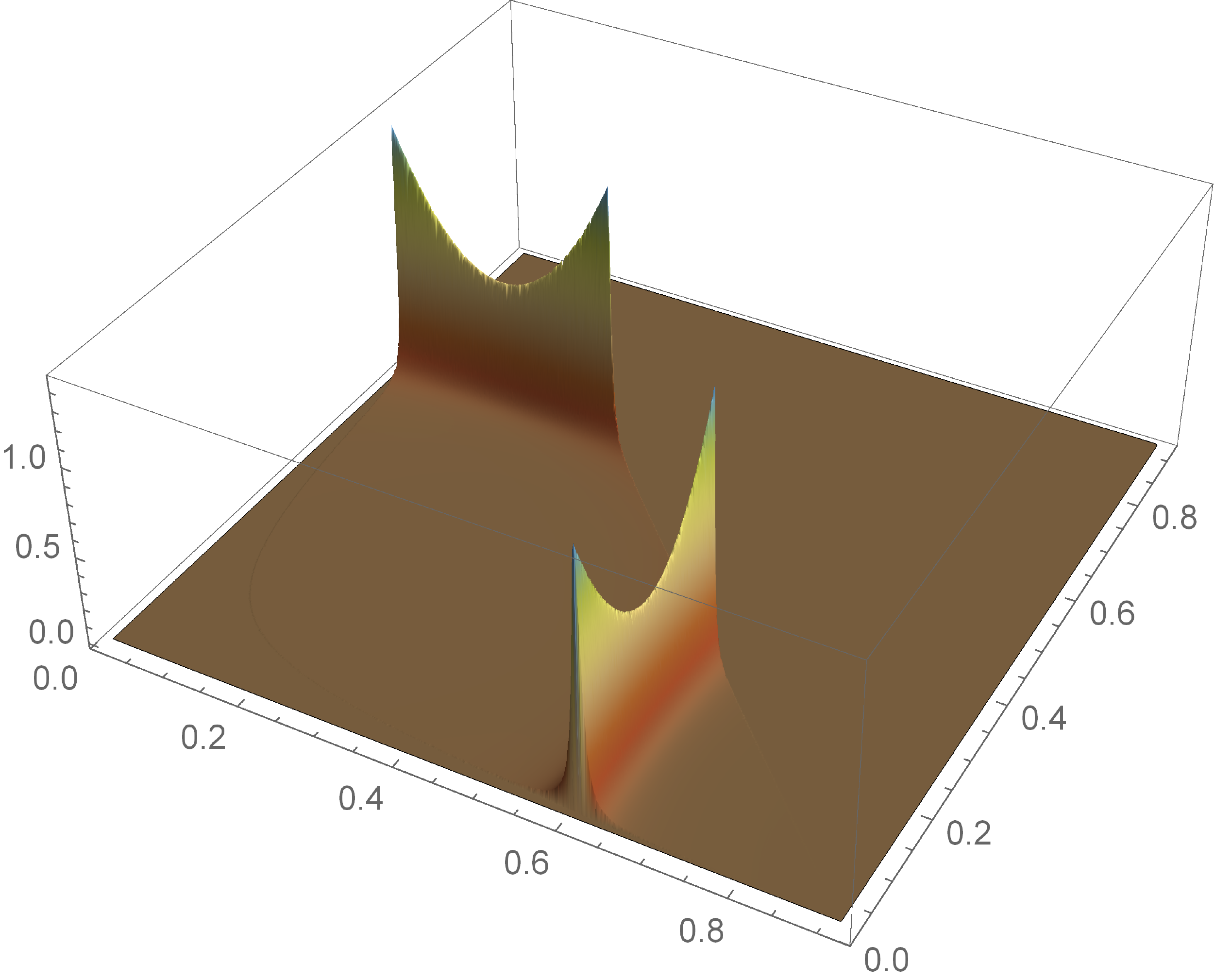}
\caption{Contribution of $\omega$-meson only.}
\label{OmegaONLY}
\end{figure}

In order to understand our results better, we plot only the contribution of $\omega$-meson, so only the terms $\sim c_{\omega}^2$
\ref{OmegaONLY}. 

\begin{figure}
\includegraphics[width=10cm]{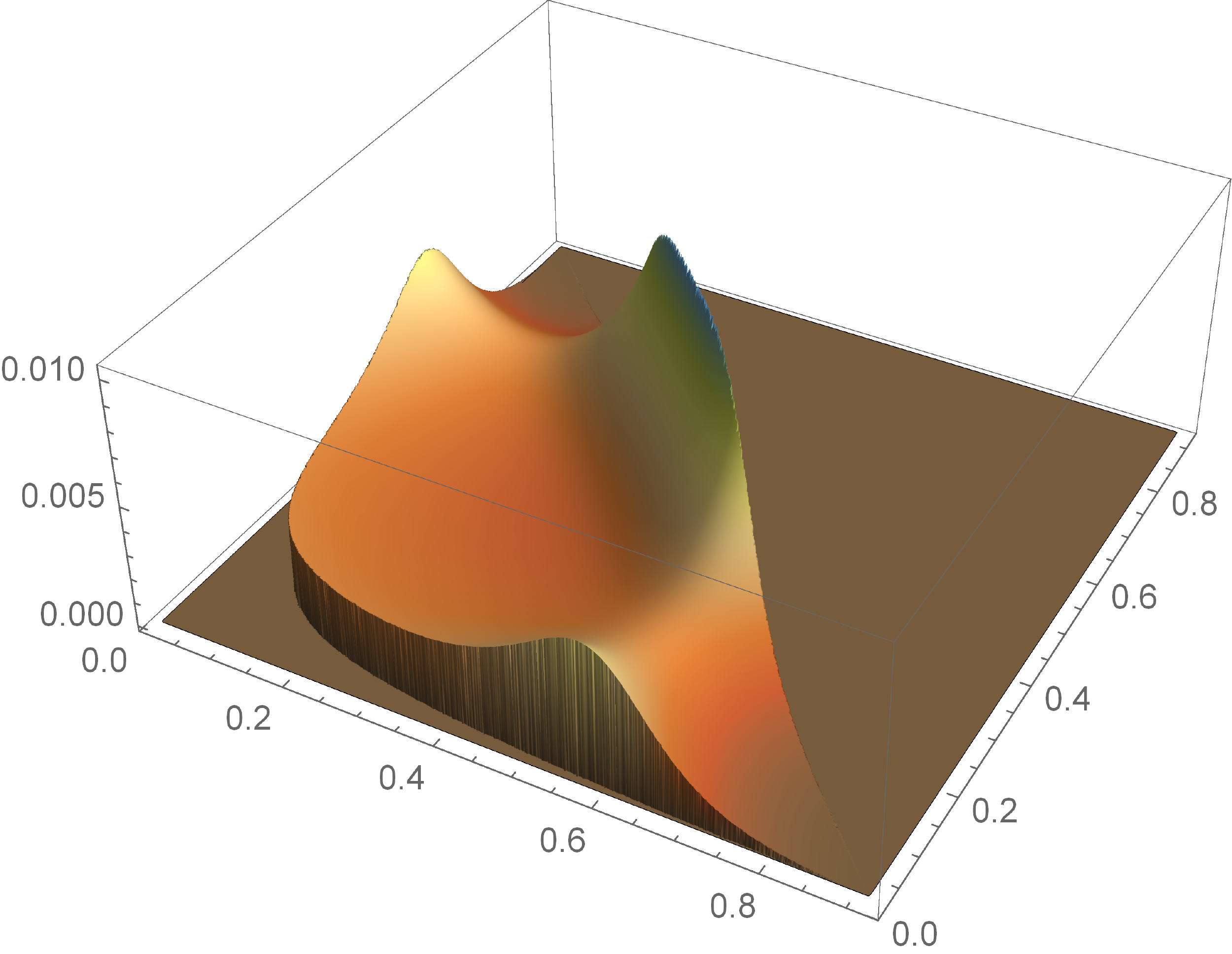}
\caption{Contribution of $\rho$-meson only.}
\label{RhoONLY}
\end{figure}

The contribution of $\rho$-meson only, so the terms $\sim c^2_{\rho}$ \ref{RhoONLY}. 

\begin{figure}
\includegraphics[width=10cm]{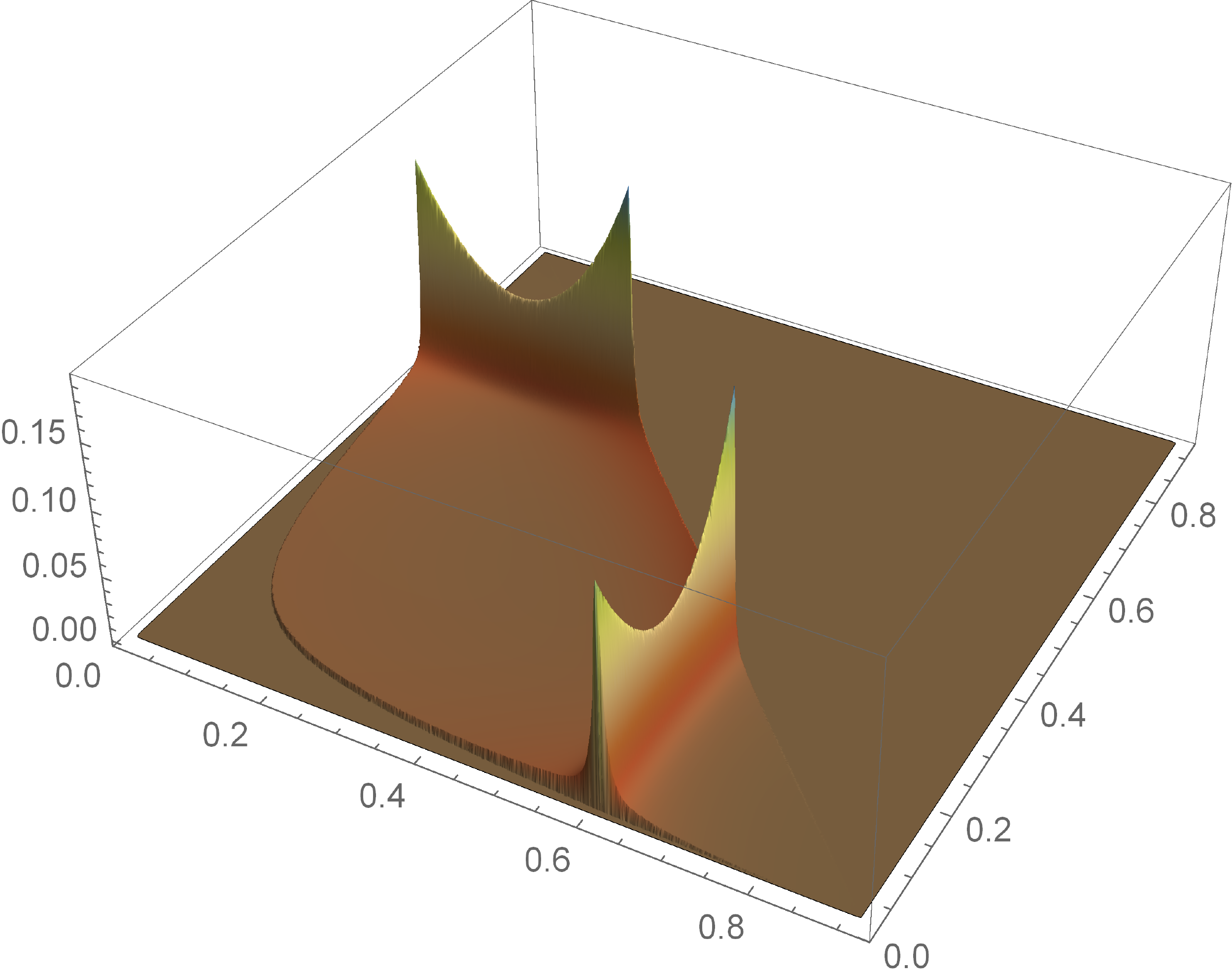}
\caption{Contribution of $\omega-\rho$ interference.}
\label{OmegaRho}
\end{figure}

And finally, the contribution of $\omega-\rho$ interference which is $\sim c_{\omega}\cdot c_{\rho}$ \ref{OmegaRho}.

If we split the decay width in the following way: $\Gamma_{total}=\Gamma_{\omega}+\Gamma_{\rho}+\underbrace{\Gamma_{\omega-\rho}}_{Interference \ \omega - \rho}$

Their relative contributions are the following:

$$\frac{\Gamma_{\omega}}{\Gamma_{total}}\approx 75 \%$$

$$\frac{\Gamma_{\rho}}{\Gamma_{total}}\approx 5 \%$$

$$\frac{\Gamma_{\omega-\rho}}{\Gamma_{total}}\approx 20 \%$$

As we see, the interference term is crucial in the area of a Dalitz plot outside the range of $\omega$  meson. 

\section{Comparison with the experiment}

In \cite{BES-III} the branching ratio was reported to be:

\begin{equation}
\mathcal{B}(\eta'\rightarrow\gamma\gamma\pi^0)_{Incl.}=(3.20 \pm 0.07 (stat) \pm 0.23(sys))\times 10^{-3}
\end{equation}

Taking the value of the total width of $\eta'$ from PDG \cite{pdg} we find the decay width:

\begin{equation}
\Gamma(\eta'\rightarrow\pi^0\gamma\gamma)=0.199 MeV \times 3.2 \times 10^{-3}  \approx 0.64 keV
\end{equation}

Clearly, we see a tension between the observed value and our theoretical result.

\section{Acknoledgements}
Y.B. very appreciates the help of LI Hai-Bo, A. Likhoded,  V. Samoylenko, V. Kiselev and M. Mikhasenko.

Y.B. is also very grateful to  Prof. Z-B. Kang for a constant help and a kind support, also his work was supported by a Graduate Dean Scholar Award at UCLA.

\section{Appendix}\label{Appendix}
\begin{figure}\label{Vertex}
  \includegraphics[width=6cm]{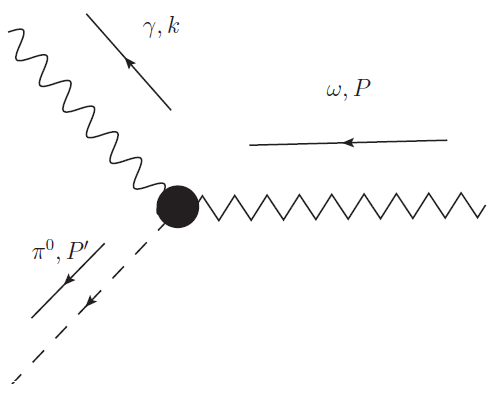}\\
  \caption{The vector-vector-pseudoscalar vertex.}
\label{fig_pig_pipi}
\end{figure}

Here we make an explicit calculation of the decay rate. We use a well-known  vector-vector-pseudoscalar vertex\ref{Vertex}. 

Take amplitude squared takes the form

$$|A|^2=G^2\epsilon_{\mu\nu\alpha\beta}\epsilon^{\mu}_{\omega}p^{\nu}\epsilon^{\alpha}_f k^{\beta}\cdot\epsilon^{\rho\sigma\tau\delta}\epsilon_{\omega\rho} p_{\sigma}\epsilon_{f\tau}k_{\delta}$$

taking into account $$\epsilon^{\alpha\beta\gamma\delta}\cdot\epsilon_{\alpha\beta\mu\nu}=
-2(\delta^{\gamma}_{\mu}\delta^{\delta}_{\nu}-\delta^{\gamma}_{\nu}\delta^{\delta}_{\mu})$$

Assuming that $\omega$-meson is at rest, we obtain, 
 $$|A|^2=2G^2(p\cdot q)^2=2G^2m^2_{\omega}q^2$$

 $$d\Gamma=\frac{1}{3}\cdot\frac{1}{2m_w}|A|^2 \frac{d^3 p_{\pi}}{(2\pi)^3 2E_{\pi}}\frac{d^3 q}{(2\pi)^3 2q}(2\pi)^4\delta^4({P-\sum p}) $$
 
Calculate $\delta$-function:
 $$ m^2_{\omega}-2 q\cdot m_{\omega}+q^2=q^2+m^2_{\pi}
 \Rightarrow q_0=\frac{m^2_{\omega}-m^2_{\pi}}{2m_{\omega}}
 $$
 
Derivative in this point is:
 $$
 1+\frac{q_0}{\sqrt{q_0^2+m^2_{\pi}}}=\frac{2m^2_{\omega}}{m^2_{\omega}+m^2_{\pi}}
 $$
 
Finally, 
  $\delta(m_{\omega}-q-\sqrt{q^2+m^2_{\pi}})= 
\frac{m^2_{\omega}+m^2_{\pi}}{2m^2_{\omega}}
\delta(q-\frac{m^2_{\omega}-m^2_{\pi}}{2m_{\omega}})$.

\begin{math}
\Gamma(\omega\rightarrow\pi^0\gamma)=\frac{1}{3}\cdot G_{|\omega\rightarrow\pi^0\gamma}^2\frac{4\pi\cdot 2m^2_{\omega}}{2m_{\omega}(2\pi)^2}\cdot
\int\frac{q^2\cdot q^2 dq}{2q\cdot 2\sqrt{q^2+m^2_{\pi}}}\delta(m_{\omega}-q-\sqrt{q^2+m^2_{\pi}})=\frac{1}{3}\cdot G_{|\omega\rightarrow\pi^0\gamma}^2 
(\frac{m_{\omega}}{4\pi})J
\end{math}

The integral equals:

\begin{math}
J=\int\frac{q^3 dq}{\sqrt{q^2+m^2_{\pi}}}\delta(m_{\omega}-q-\sqrt{q^2+m^2_{\pi}})=\frac{1}{4}\frac{(m^2_{\omega}-m^2_{\pi})^3}{m^2_{\omega}(m^2_{\omega}+m^2_{\pi})}\cdot\frac{m^2_{\omega}+m^2_{\pi}}{2m^2_{\omega}}=
\frac{(m^2_{\omega}-m^2_{\pi})^3}{8m^4_{\omega}}
\end{math}

Finally, we obtain the result for the decay $\omega\rightarrow\pi^0\gamma$

\begin{equation}
\Gamma(\omega\rightarrow\pi^0\gamma)=\frac{1}{3}\cdot G_{\omega\rightarrow\pi^0\gamma}^2 \frac{(m^2_{\omega}-m^2_{\pi})^3}{32\pi\cdot m^3_{\omega}}
\end{equation}

For the decay $\rho\rightarrow\pi^0\gamma$ we just have to replace $\omega\rightarrow\rho$. 

For the decays $\eta'\rightarrow\omega\gamma$ and  $\eta'\rightarrow\rho\gamma$ we have to replace $\omega\rightarrow\eta'$,  $\pi^0\rightarrow\omega$ and  $\pi^0\rightarrow\rho$ for each of decay respectively. Also there is no $\frac{1}{3}$ factor since $\eta'$ is a pseudoscalar particle.

\end{document}